\documentclass[aps,reprint,superscriptaddress,pra,%
preprintnumbers,amsmath,amssymb,letterpaper,floatfix,showpacs]{revtex4-1}
\usepackage[colorlinks,linkcolor = red,anchorcolor = blue,%
urlcolor = blue, citecolor = blue]{hyperref}
\usepackage{graphicx}
\usepackage{times}
\usepackage{dcolumn}
\usepackage{color}
\usepackage{url}

\begin{document}

\preprint{}

\title{\boldmath Renormalizations in unconventional superconducting states of Ce$_{1-x}$Yb$_{x}$CoIn$_{5}$}

\author{Z. F. Ding}
\author{J. Zhang}
\author{C. Tan}
\affiliation{State Key Laboratory of Surface Physics, Department of Physics, Fudan University, Shanghai 200433, China}
\author{K. Huang}
\altaffiliation{Current address: Lawrence Livermore National Laboratory, Livermore, California 94550, USA.}
\affiliation{State Key Laboratory of Surface Physics, Department of Physics, Fudan University, Shanghai 200433, China}
\author{I. Lum}
\affiliation{Department of Physics, University of California, San Diego, La Jolla, California 92093,USA}
\author{O. O. Bernal}
\affiliation{Department of Physics and Astronomy, California State University, Los Angeles, California 90032, USA}
\author{P.-C. Ho}
\affiliation{Department of Physics, California State University, Fresno, California 93740,USA}
\author{D. E. MacLaughlin}
\affiliation{Department of Physics and Astronomy, University of California, Riverside, California 92521, USA}
\author{M. B. Maple}
\affiliation{Department of Physics, University of California, San Diego, La Jolla, California 92093,USA}
\author{L. Shu}
\altaffiliation[Corresponding Author: ]{leishu@fudan.edu.cn.}
\affiliation{State Key Laboratory of Surface Physics, Department of Physics, Fudan University, Shanghai 200433, China}
\affiliation{Collaborative Innovation Center of Advanced Microstructures, Nanjing 210093, China}
\date{\today}

\begin{abstract}
We have measured the superconducting penetration depth~$\Lambda(T)$ in the heavy-fermion/intermediate-valent superconducting alloy series~Ce$_{1-x}$Yb$_x$CoIn$_5$ using transverse-field muon spin relaxation, to study the effect of intermediate-valent Yb doping on Fermi-liquid renormalization. From $\Lambda(T)$ we determine the superfluid density $\rho_s(T)$, and find that it decreases continuously with increasing nominal Yb concentration~$x$, i.e., with increasing intermediate valence. The temperature-dependent renormalization of the ``normal'' fluid density~$\rho_N(T) = \rho_s(0) - \rho_s(T)$ in both the heavy-fermion and intermediate valence limits is proportional to the temperature-dependent renormalization of the specific heat. This indicates that the temperature-dependent Fermi-liquid Landau parameters of the superconducting quasiparticles entering the two different physical quantities are the same. These results represent an important advance in understanding of both intermediate valence and heavy-fermion phenomena in superconductors.

\end{abstract}


\maketitle

\section{INTRODUCTION}

It is remarkable that Landau Fermi-liquid theory \cite{Landau56,*Landau58} applies to heavy-fermion systems where mass enhancements due to many body effects often exceed $O(10^2)$. It is understood that the reason for the renormalizations is the Kondo effect at each of the magnetic ions. This is described by the periodic Anderson model~\cite{VarmaYafet}, in which one of the charge states of the magnetic ions (usually $4f$ or $5f$) is well below the Fermi energy and another charge state is well above, being separated from the former by the local repulsion parameter~$U$. In contrast, in intermediate valence compounds two charge states are close to the Fermi energy, so that magnetic moment and charge fluctuations are equally important. Intermediate valence is therefore an even more subtle problem than Kondo-effect-induced heavy-fermion behavior.

Novel properties like quantum criticality, non-Fermi-liquid behavior, and unconventional superconductivity originate in such systems due to the strong correlation between localized and itinerant electrons. Research on heavy-fermion superconductors can also help us to understand the pairing mechanism of unconventional superconductivity. It is important, therefore, to fully characterize the many-body effects in heavy-fermion and intermediate-valence compounds. The alloy system~Ce$_{1-x}$Yb$_{x}$CoIn$_{5}$~\cite{Shu11,Booth11} offers a very good opportunity to do this.

CeCoIn$_{5}$ has one of the highest superconducting transition temperatures (2.3~K) among heavy-fermion superconductors~\cite{Petrovic01a}. Quantum criticality is observed when the system is tuned by either pressure~\cite{Sidorov02} or magnetic field~\cite{Paglione03,Bianchi03b,Tanatar07,Hu12}. Investigation of the effect of Yb substitution for Ce on the structure and physical properties of CeCoIn$_5$ has been motivated by the electron-hole analogy between the Ce$^{3+}$ ($4f^1$) and Yb$^{3+}$ (4$f^{13}$) electronic configurations, together with the unstable valence~$\nu$ of Ce ($3{+}\leqslant$ $\nu_\mathrm{Ce}$ $\leqslant 4{+}$) and Yb ($2{+} \leqslant$ $\nu_\mathrm{Yb}$ $\leqslant 3{+}$)~\cite{Shu11,Booth11}.

A number of remarkable phenomena have been observed in Ce$_{1-x}$Yb$_{x}$CoIn$_5$ throughout the entire range of nominal Yb composition~$x$~\cite{Shu11, Booth11, Mizukami11, White12, Shimozawa12, Polyakov12, Dudy13, Hu13, Jang14, Kim15, Erten15, Zhang15, Xu16}. The $T$-$x$ phase diagram is unconventional for heavy-fermion superconductors, in which the superconducting transition temperature~$T_c$ and the Kondo coherence temperature~$T_\mathrm{coh}$ do not track each other~\cite{Shu11}. Both $T_c$ and the $T = 0$ electronic specific heat coefficient~$\gamma_0$ are systematically suppressed with Yb doping~\cite{Shu11,Booth11}, but the suppression of $T_c$ is much less than for other rare-earth substitutions\cite{Petrovic02,Hu08}. While the Ce valence is near $3{+}$ for all $x$, the Yb valence is ${\sim}2.3+$ for $x \geqslant 0.3$ and increases rapidly with decreasing $x \leqslant 0.2$ towards $3+$ as $x \to 0$~\cite{Booth11,Dudy13}.

We report results of muon spin relaxation ($\mu$SR) and specific heat measurements in Ce$_{1-x}$Yb$_{x}$CoIn$_{5}$ alloys for $0 \leqslant x \leqslant 0.5$, and discuss their implications for the theory of heavy-fermion and intermediate-valence systems. The $\mu$SR data yield the magnetic penetration depth~$\Lambda(T)$, from which the superfluid density $\rho_s$ is obtained. With increasing $x$, $\rho_s$ initially decreases, as expected from the transition from heavy-fermion character to intermediate valence~\cite{Varma76}, and then remains constant for $x>0.2$ where the valence of Yb does not change. In agreement with a recent theory~\cite{Miyake18}, the renormalized temperature dependence of the superfluid density is similar to that of the specific heat, indicating that the temperature-dependent Fermi-liquid Landau parameters entering these two different physical quantities are the same.

Transverse-field $\mu$SR (TF-$\mu$SR)~\footnote{I.e., $\mu$SR in a field applied perpendicular to the initial muon spin direction.} has been proved to be an effective probe of the magnetic field distribution in the vortex state of type II superconductors~\cite{Sonier00,Sonier07}. In a magnetic field~$B$, a muon spin precesses at the Larmor frequency~$\omega = \gamma_\mu B$, where $\gamma_\mu = 2\pi{\times}135.5342$~MHz/T is the muon gyromagnetic ratio, before decaying with a lifetime $\approx 2.2\ \mu$s into a positron and two neutrinos. The positron is emitted preferentially along the direction of the muon spin due to parity violation. The time evolution of the muon spin polarization is determined by detecting decay positrons from an ensemble of ${\sim}2 \times 10^7$ muon events.

Under certain conditions of field and temperature~\cite{Sonier00,Sonier07}, the muon spin relaxation rate in the vortex state is given by $\gamma_\mu \delta B_\mathrm{rms}$, where $\delta B_\mathrm{rms} = \langle \delta B^2\rangle^{1/2}$ is the rms width of the field distribution in the flux-line lattice (FLL)\@. In the extreme Type-II or London limit ($\Lambda \gg \text{coherence length } \xi$)
\begin{equation} \label{eq:Brandt}
\langle\delta B^2\rangle = 0.00371 \, \Phi_0^2/\Lambda^4\,,
\end{equation}
where $\Phi_0$ is the flux quantum~\cite{Brandt88}. In turn, from the London equations $\Lambda$ is related to $\rho_s$ and the carrier effective mass~$m_\mathrm{eff}$ by
\begin{equation} \label{eq:lambda1}
\frac{1}{\Lambda^{2}} = \frac{4\pi e^2 \rho_s}{m_\mathrm{eff}c^2}\,.
\end{equation}
Thus TF-$\mu$SR experiments yield information on $\rho_s(T)$.

\section{EXPERIMENT}

High quality single crystals of Ce$_{1-x}$Yb$_{x}$CoIn$_5$, $x = 0, 0.05, 0.125, 0.2, 0.3, 0.4$, and $0.5$, were synthesized using an indium self-flux method~\cite{Zapf01}. After centrifuging and etching in HCl solution to remove excess indium, large thin plate-like single crystals were obtained. Crystal structures were verified by x-ray diffraction. Specific heat measurements were made down to 50~mK using a Quantum Design Physical Property Measurement System platform equipped with a dilution refrigerator. Transition temperatures~$T_c$ from these data were consistent with values reported in Refs.~\onlinecite{Shu11,Booth11}, confirming the Yb doping concentrations. Single crystals with flat $ab$ planes were selected, aligned, and glued onto a silver plate holder using dilute GE varnish, covering a $10{\times}10$~mm$^2$ area. $\mu$SR experiments were carried out using the M15 beam line at TRIUMF, Vancouver, Canada, in a top-loading dilution refrigerator with a base temperature of 20~mK\@. Spin-polarized positive muons were implanted into a sample in a transverse external magnetic field $\mu_0 H = 30$~mT [$H \ll H_{c2}(0)$]. The sample was field cooled into the superconducting state from above $T_c$.

\begin{figure}[ht]
 \includegraphics[width = 0.45\textwidth]{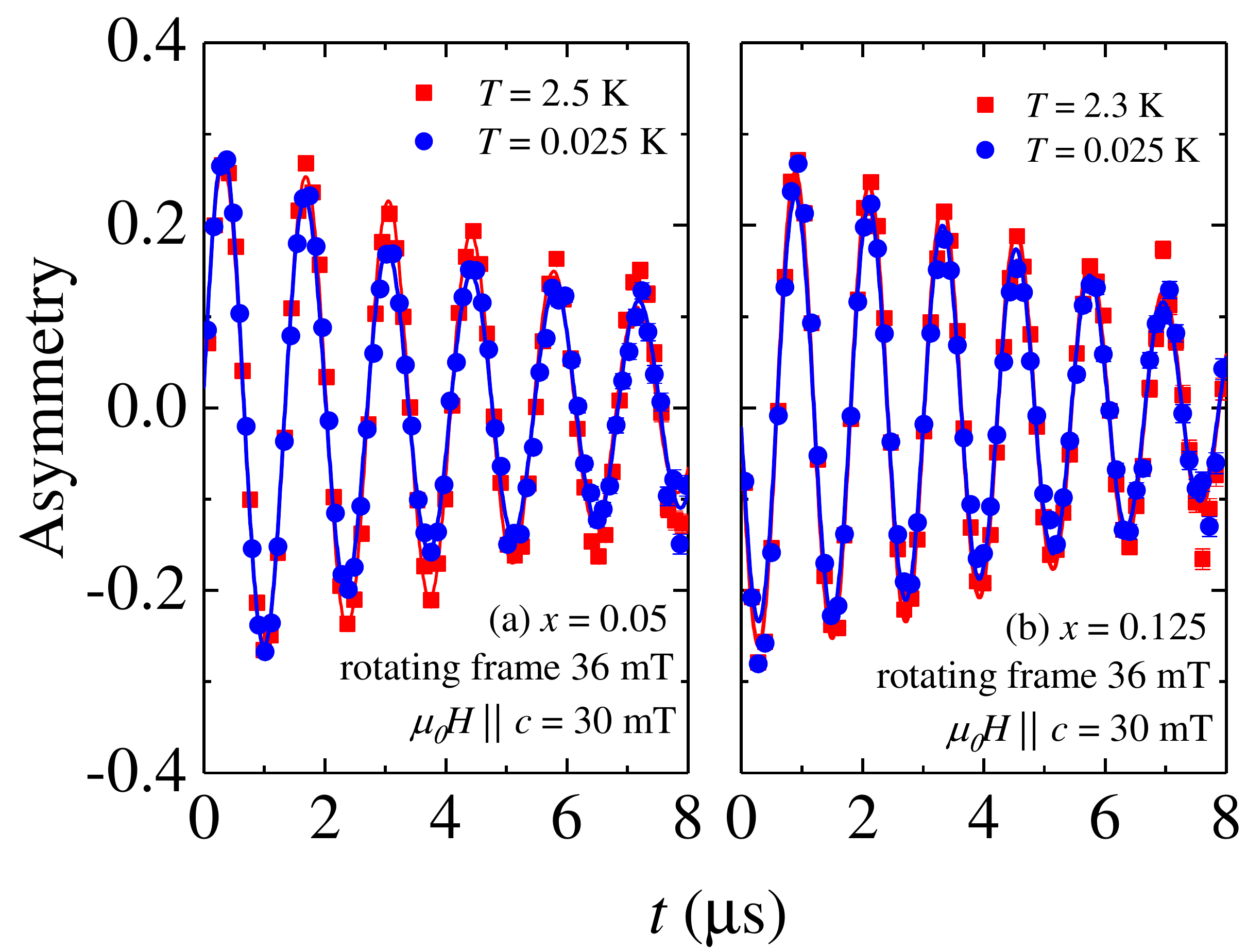}
 \caption{(Color online) $\mu$SR asymmetry spectra $A(t)$ from Ce$_{1-x}$Yb$_{x}$CoIn$_{5}$, $\mu_0H =  30$ mT, $H \parallel c$. (a)~$x = 0.05$. (b)~$x = 0.125$. Red squares: normal state. Blue circles: superconducting state. Solid curves: fits to the data (see text).}
 \label{fig:Asy}
\end{figure}

Typical TF-$\mu$SR asymmetry spectra from the normal and superconducting states of Ce$_{1-x}$Yb$_{x}$CoIn$_5$ are shown in Fig.~\ref{fig:Asy} for $x = 0.05$ and 0.125.
Each spectrum consists of two contributions: a signal from muons stopping in the sample, and a background signal from muons that miss the sample and stop in the silver sample holder. In the superconducting state the damping of the sample signal is enhanced due to the field broadening generated by the FLL\@. The enhancement is smaller for the $x = 0.125$ sample, due to the depression of $\rho_s$ by Yb ions.

The spectra from both undoped CeCoIn$_5$~\cite{Shu14} and the Yb-doped alloys are well fit using the function
\begin{equation}
 \label{eq:Asy}
 \begin{split}
 A(t) = &\ A_0\bigl[f_s\exp(-\textstyle{\frac{1}{2}}\sigma_s^2t^2)\cos(\omega_s t+\phi_s)\\
 & +(1-f_s)\textstyle{\exp(-\frac{1}{2}} \sigma_b^2t^2)\cos(\omega_b t+\phi_b)\bigr] \,,
 \end{split}
 \end{equation}
where the first and second terms represent sample and background signals, respectively. Here $A_0$ is the initial asymmetry and $f_s$ is the fraction of muons that stop in the sample. The Gaussian relaxation rate $\sigma_s$ from the sample is due to nuclear dipolar fields in the normal state, and is enhanced in the superconducting state by the FLL field inhomogeneity. The precession frequency $\omega_s$ is reduced due to diamagnetic screening. The background relaxation rate $\sigma_b$ is negligibly small, and the initial phases $\phi_b$ and $\phi_s$ and the background frequency $\omega_b$ are constant. The curves in Fig.~\ref{fig:Asy} are fits to Eq.~(\ref{eq:Asy}).

\section{RESULTS}

\begin{figure}[ht]
 \begin{center}
 \includegraphics[width = 0.45\textwidth]{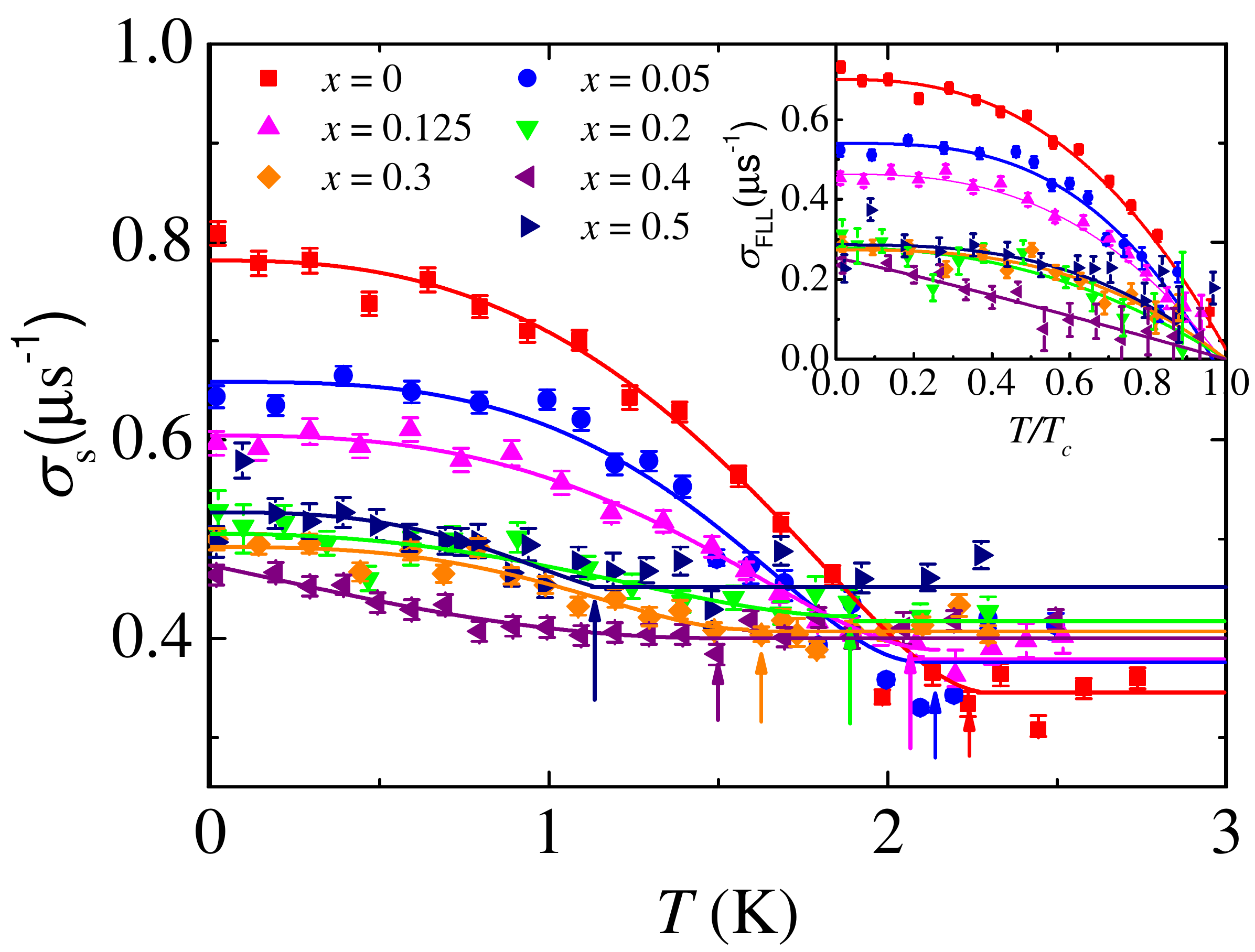}
 \caption{(Color online) Temperature dependencies of TF-$\mu$SR relaxation rate~$\sigma_s$ in Ce$_{1-x}$Yb$_{x}$CoIn$_{5}$, $\mu_0H =  30$ mT $\parallel c$. Data for $x = 0$ is from Ref.~\onlinecite{Shu14}. Solid curves: power-law fits to the data (see text). Arrows: $T_c$ from transport measurements~\cite{Shu11,Booth11}. Inset: flux-line lattice relaxation rate~$\sigma_{\rm FLL}$ from Eq.~(\protect\ref{eq:sigma}) vs normalized temperature $T/T_c$.}
 \label{fig:Rlx}
 \end{center}
\end{figure}

Figure~\ref{fig:Rlx} shows the temperature dependencies of $\sigma_s$ in Ce$_{1-x}$Yb$_{x}$CoIn$_{5}$ for $x = 0$ (data from Ref.~\onlinecite{Shu14}), 0.05, 0.125, 0.2, 0.3, 0.4, and 0.5. The arrows indicate $T_c$ determined from transport measurements~\cite{Shu11,Booth11}.
The common features of these data are (1)~the temperature independence of $\sigma_s$ above $T_c$, and (2)~a clear increase of $\sigma_s$ below $T_c$ due to the formation of the FLL.

The internal field distribution in the vortex state is the convolution of the field distribution due to the vortex lattice and the nuclear dipolar field distribution of the host material:
 \begin{equation}
 \label{eq:sigma}
 \sigma_s^2 = \sigma_{\rm FLL}^2+\sigma_{\rm dip}^2,
 \end{equation}
where $\sigma_{\rm dip}^2$ is temperature independent. Values of $\sigma_{\rm dip}$ for all $x$ are given in Table~\ref{table-A}. With increasing $x$, $\sigma_{\rm dip}$ initially increases slightly, due to a contribution from Yb nuclear dipolar fields. The small reduction of $\sigma_{\rm dip}$ for $x > 0.2$ might be due to a decrease in the contact (RKKY) interaction, since the Fermi surface starts to change at this Yb concentration~\cite{Polyakov12}.

In the inset to Fig.~\ref{fig:Rlx}, $\sigma_{\rm FLL}(T)$ from Eq.~(\ref{eq:sigma}) is plotted vs reduced temperature $T/T_c$ for all $x$. The data can be fit with the power law
\begin{equation}
 \label{eq:sigmaFLL}
 \sigma_{\rm FLL}(T) = \sigma_{\rm FLL}(0)\left[1-(T/T_c)^{n_\sigma}\right]\,, \quad T < T_c\,,
 \end{equation}
with best-fit values of $\sigma_{\rm FLL}(0)$ and $n_\sigma$ that are listed in Table~\ref{table-A}.
\begin{table*}[t]
\caption{Parameters from fits to TF-$\mu$SR and specific-heat data from Ce$_{1-x}$Yb$_{x}$CoIn$_{5}$. Muon nuclear dipolar relaxation rate~$\sigma_\mathrm{dip}$, $T{=}0$ muon FLL relaxation rate~$\sigma_\mathrm{FLL}(0)$, transition temperature~$T_c$, and power-law exponent~$n_\sigma$ from fits of Eq.~(\ref{eq:sigmaFLL}) to $\sigma_{\rm FLL}(T)$. The $T_c$ values for $x\geq0.2$ were fixed at values determined from heat capacity measurements. Zero-temperature penetration depth~$\Lambda(0)$ from Eq.~(\ref{eq:Brandt}). Specific-heat exponent~$n_\gamma$: from fits of power law~$C_{e} /T =  \gamma_0 +aT^{\,n_\gamma}$ to specific heat data.}\label{table-A}
\begin{ruledtabular}
\begin{tabular}{lccccccc}
$x$ & 0\footnote{\,$\mu$SR data from Ref.~\protect\onlinecite{Shu14}.} & 0.05 & 0.125 & 0.2 & 0.3 & 0.4 & 0.5\\
\hline
 $\sigma_{\rm dip}\ (\mu\text{s}^{-1})$ & 0.316(1) & 0.346(4) & 0.415(3) & 0.426(3) & 0.407(3) & 0.390(3) & 0.449(5)\\
 $\sigma_{\rm FLL}(0)\ (\mu\text{s}^{-1})$ & 0.67(1) & 0.55(1)& 0.46(1)& 0.28(2) & 0.27(1) & 0.24(2)& 0.28(2)\\
 $T_c$ (K) & 2.27(2) & 2.18(3) & 2.14(4) & 1.90 & 1.59 & 1.50 & 1.3\\
 $n_\sigma$ & 2.4(2) & 3.1(2) & 2.6(3) & 2.3(6) & 2.8(4) & 1.7(4) & 2.8(6)\\
 $\Lambda (0)$ ($\mu$m) & 0.386(3) & 0.441(4) & 0.483(5) & 0.62(2) & 0.63(1) & 0.67(3) & 0.61(2) \\
 $n_{\gamma}$\footnote{Specific heat data from Ref.~\onlinecite{Shu11} and the present measurements.} & 2.20(5) & 2.54(1) & 2.69(2) & &2.49(5) & & \\
\end{tabular}
\end{ruledtabular}
\end{table*}
The values of $T_c$ are consistent with results of transport measurements~\cite{Shu11,Booth11}. The zero-temperature penetration depths~$\Lambda(0)$ obtained from Eq.~(\ref{eq:Brandt}) are also listed in Table~\ref{table-A}. Previous results~\cite{DeBeer-Schmitt06,Xu16} yield coherence lengths in Ce$_{1-x}$Yb$_{x}$CoIn$_5$ shorter than 82~\AA\@. Comparison with penetration depths from Table~\ref{table-A} shows that the entire alloy system is in the extreme Type-II region, as is necessary for the applicability of Eq.~(\ref{eq:Brandt}).

\begin{figure}[ht]
 \begin{center}
 \includegraphics[width = 0.45\textwidth]{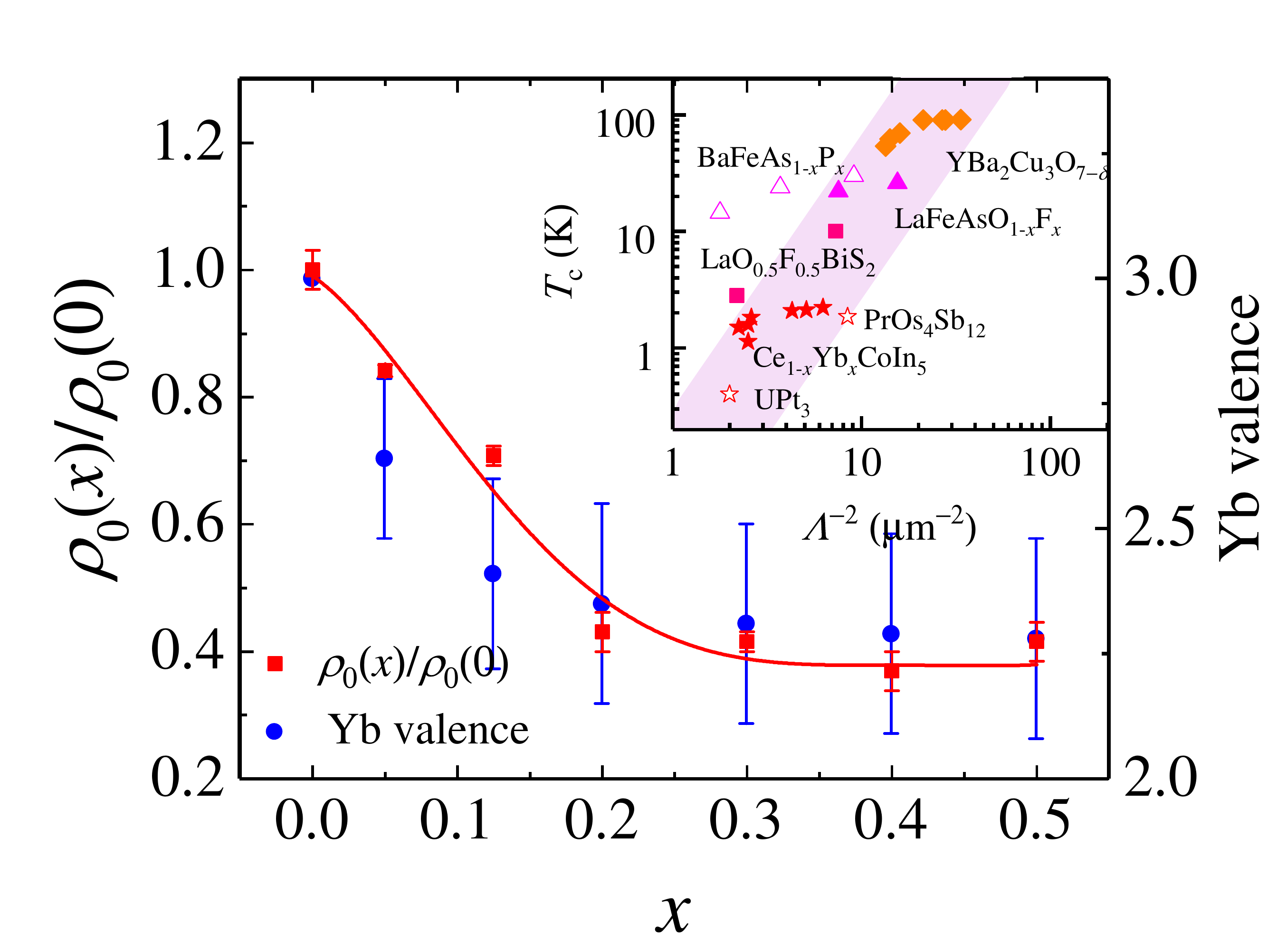}
 \caption{Dependence of normalized $T{=}0$ superfluid densities~$\rho_0(x)/\rho_0(0)$ and Yb valence~\cite{Booth11,Dudy13} on nominal Yb concentration~$x$ in Ce$_{1-x}$Yb$_{x}$CoIn$_{5}$. The curve is a guide to the eye. Inset: Uemura plot of $T_c$ vs $\Lambda^{-2}$ for various superconductors~\cite{Broholm90,Luetkens08,MacLaughlin02,%
Hashimoto12,Zhang16,Lamura13} including Ce$_{1-x}$Yb$_{x}$CoIn$_{5}$.}
 \label{fig:rho}
 \end{center}
\end{figure}

The normalized zero-temperature superfluid densities~$\rho_0(x)/\rho_0(0) = \Lambda_0^2(0)/\Lambda_0^2(x)$ in Ce$_{1-x}$Yb$_{x}$CoIn$_5$ are plotted vs $x$ in Fig.~\ref{fig:rho} (the subscript ``0'' signifies zero temperature).
It can be seen that $\rho_0(x)$ decreases continuously with increasing $x$; the decrease is rapid up to $x = 0.2$, and saturates for higher $x$. As noted above, the Yb valence~$\nu_\mathrm{Yb}$ also decreases rapidly with increasing $x$, from nearly $3+$ near $x = 0$ to ${\sim} 2.3+$ above $x \approx 0.2$~\cite{Booth11,Dudy13}. The consistent change of superfluid density and Yb valence correlates well with de Haas-van Alphen results~\cite{Polyakov12} showing a smooth change of Fermi surface up to $x = 0.2$ and a drastic reconstruction above $x = 0.55$.

\begin{figure}[ht]
 \begin{center}
 \includegraphics[width = 0.45\textwidth]{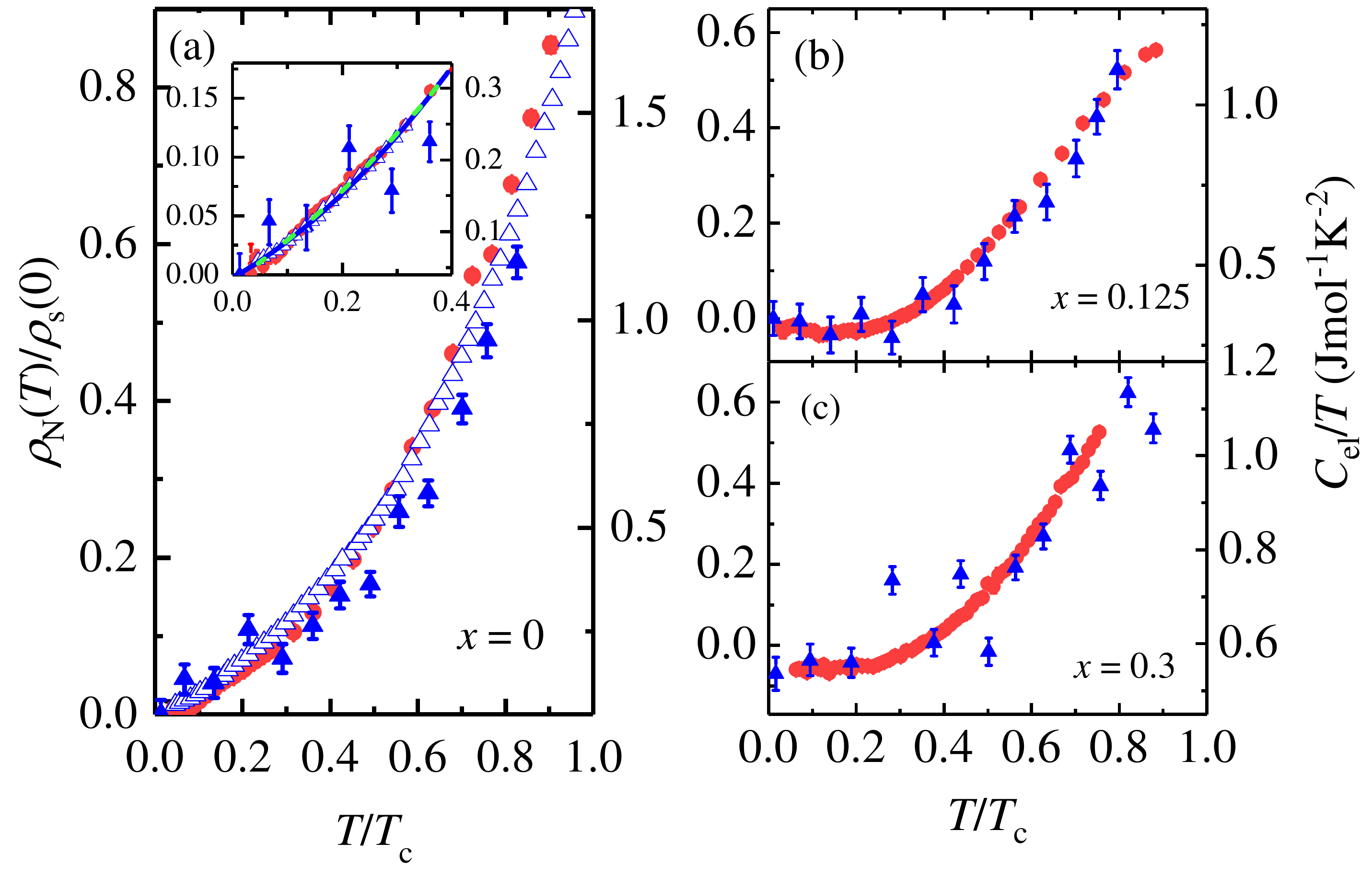}
 \caption{Temperature dependencies of normalized ``normal'' fluid density~$\rho_N(T)/\rho_s(0) = 1 - \Lambda^2(0)/\Lambda^2(T)$ and electronic specific heat coefficient~$C_\mathrm{el}/T$ in superconducting Ce$_{1-x}$Yb$_{x}$CoIn$_{5}$. Solid blue triangles: $\rho_N$ measured by $\mu$SR experiment. Open blue triangles: $\rho_N$ measured using the TDO technique~\cite{Hashimoto13}. Red circles: $C_\mathrm{el}/T$ (Ref.~\onlinecite{Shu11} and present measurements). (a)~$x = 0$. Inset: low-temperature data. Curves: fits of Eqs.~(\ref{eq:rho_N}) and (\ref{eq:Cel}) to the data, and the fitting curve of specific heat data is plotted in light green dashed curve for comparison. (b)~$x = 0.125$. (c)~$x = 0.3$.}
 \label{fig:rho_C}
 \end{center}
\end{figure}

The temperature dependence of the so-called ``normal-fluid'' density~$\rho_{N}(T) = \rho_s(0) -\rho_s(T)$ of thermal excitations from the ground state is shown in Fig.~\ref{fig:rho_C} for $x = 0$, 0.125, and 0.3.
The data are normalized to $\rho_s(0)$, so that from Eq.~(\ref{eq:lambda1}) $\rho_N(T)/\rho_s(0) = 1 - \Lambda^2(0)/\Lambda^2(T)$. For $x = 0$, $\rho_N(T/T_c)$ derived from tunnel diode oscillator (TDO) penetration depth measurements~\cite{[{}]  [{. In TDO measurements the change of penetration depth~$\Lambda(T) - \Lambda(0)$ is obtained from the inductance of a coil wrapped around the sample. In Fig.~\ref{fig:rho_C}(a) $\rho_N(T)/\rho_s(0) = 1 - \Lambda^2(0)/\Lambda^2(T)$ is calculated using the $\mu$SR result~$\Lambda(0)=0.386~\mu$m.}] Hashimoto13} is plotted in Fig.~\ref{fig:rho_C}(a) in addition to $\mu$SR results~\cite{[{See also }]Chia03, *Ormeno02, *Ozcan03}. The consistency between the two different techniques suggests the reliability of the measurements.

The electronic specific heat~$C_\mathrm{el}(T)$ was determined by subtracting the Schottky contribution~$C_\mathrm{Sch}(T) \propto T^{-2}$ from quadrupole-split $^{115}$In nuclear spins from the measured values. The temperature dependences of the specific heat coefficient~$C_\mathrm{el}(T)/T$ for $0 < T \leqslant T_c$~\footnote{Ref.~\onlinecite{Shu11} and the present measurements.} are also shown in Fig.~\ref{fig:rho_C}. The data were fit using the power law~$C_\mathrm{el} /T =  \gamma_0 +aT^{\,n_\gamma}$, yielding the corresponding exponent~$n_\gamma$ given in Table~\ref{table-A}.

\section{DISCUSSION}

In heavy-fermion compounds the heavy electrons arise from the renormalization of $f$ moments to itinerant electrons by the Kondo effect~\cite{Varma85}, through exchange interactions with the $s$, $p$, and $d$ conduction bands. By doping Yb in CeCoIn$_5$, the Kondo-effect-derived heavy fermions develop into intermediate-valence states, which are expected to have much less mass enhancement than in the Kondo limit. However, for $x \gtrsim 0.2$, where the Yb valence is ${\sim} 2.3+$, substantial normal-state mass enhancement is still observed in the specific heat and resistivity of Ce$_{1-x}$Yb$_{x}$CoIn$_5$ up to $x \approx 0.65$~\cite{Shu11,Booth11,Polyakov12,Dudy13}. This suggests that strong correlated-electron effects still exit in the intermediate-valence state where the superfluid density is suppressed. An Uemura plot~\cite{Uemura89} is shown as an inset in Fig.~\ref{fig:rho}, including Ce$_{1-x}$Yb$_{x}$CoIn$_{5}$ and other unconventional superconductors such as UPt$_3$ and PrOs$_4$Sb$_{12}$. Although the behavior within the alloy series itself is not linear, all the Ce$_{1-x}$Yb$_{x}$CoIn$_{5}$ alloys fall in the shaded region, indicating unconventional superconductivity in both heavy-fermion and intermediate-valence limits,

As shown in Table~\ref{table-A}, rough agreement is found between the power-law exponents~$n_\sigma$ and $n_\gamma$, suggesting similar renormalizations of the superfluid density and the superconducting-state specific heat in Ce$_{1-x}$Yb$_{x}$CoIn$_{5}$. A recent extension of Landau theory~\cite{Miyake18} provides an understanding of  thermodynamic properties and response functions in the superconducting state of a singular Fermi liquid, and in particular accounts for this commonality. The microscopic basis of this theory is the separation of the single-particle Green's function $G(\mathbf{k}, \omega)$ into a coherent (or quasiparticle) part with poles and an incoherent analytic part. The theory yields correlation functions of conserved quantities for which the incoherent part is shown to give no contribution. For Ce$_{1-x}$Yb$_{x}$CoIn$_{5}$, non-Fermi-liquid behavior has been observed and the system is near the quantum critical fluctuation regime~\cite{Hu13}, where $G(\mathbf{k}, \omega)$ has branch cuts. Landau theory cannot be applied to such singular Fermi liquids. However, a form of Landau theory may still be applied to the superconducting state that arises from them, because the density of low-energy excitations in the superconducting state tends to zero (in the pure limit) or is analytic. The theory yields a common renormalization of the specific heat and the superfluid density, and hence predicts the same temperature dependencies of these two properties in the superconducting state.

We explore these temperature dependencies further by examining the low-temperature temperature deviations from zero-temperature values (Figure~\ref{fig:rho_C}) using expansions in powers of $T/T_c$~\footnote{This comparison is less {\it ad hoc} than the empirical assumption of power-law behavior over the entire superconducting temperature range discussed previously.}. For $x = 0$ fits of the expansions
\begin{equation}
 \label{eq:rho_N}
\rho_N(T)/\rho_s(0) = g_1^{\rho_N}(T/T_c)+g_2^{\rho_N}(T/T_c)^2 + \text{const.}
 \end{equation}
and
 \begin{equation}
 \label{eq:Cel}
C_\mathrm{el}/T = g_1^{C_\mathrm{el}}(T/T_c)+g_2^{C_\mathrm{el}}(T/T_c)^2 + \text{const.}
 \end{equation}
to the data of Fig.~\ref{fig:rho_C}(a) for $T/T_c\,{\leqslant}\,0.4$ yield $g_1^{\rho_N} = 0.28$, $g_2^{\rho_N} = 0.44$, $g_1^{C_\mathrm{el}} =  0.46~\text{J/mol K}^2$, and $g_2^{C_\mathrm{el}} = 0.71~\text{J/mol K}^2$. The fit curves for $\rho_N(T)/\rho_s(0)$ and $C_\mathrm{el}(T)/T$ in Fig.~\ref{fig:rho_C}(a) are almost identical and, correspondingly, the ratios~$g_1/g_2$ are about the same (0.64 and 0.65, respectively). This is evidence that the same expansion for the temperature dependence of the renormalization describes both these properties in CeCoIn$_5$. No TDO data have been reported For $x = 0.125$ and 0.3, but a few $\rho_N(T)$ points are available from $\mu$SR\@. Here a rough correspondence between $\rho_N$ and $C_\mathrm{el}/T$ is also found [Figs.~\ref{fig:rho_C}(b) and (c)]. This suggests that renormalization of $C_\mathrm{el}/T$ and $\rho_N$ in the superconducting states of Ce$_{1-x}$Yb$_{x}$CoIn$_{5}$ is the same for both heavy fermion and intermediate valence states. For all concentrations the scaling factor~$[C_\mathrm{el}(T)/T]/[\rho_N(T)/\rho_s(0)]$ decreases monotonically with Yb concentration, roughly tracking the normal-state specific heat coefficient. The residual specific heat in the zero-temperature limit may be due to an impurity band that forms in line nodes in the energy gap, as suggested previously~\cite{Shu11, Movshovich01}.

In conclusion, the superfluid densities of Yb-substituted alloys of the Kondo heavy-fermion compound CeCoIn$_5$ have been measured using $\mu$SR\@. We find that the superfluid density decreases continuously with increasing Yb substitution $x$. The decrease is rapid up to $x = 0.2$, and saturates for higher $x$ where intermediate valence becomes important. In both the heavy-fermion and intermediate-valence limits the temperature-dependent renormalization of $\rho_N$ is proportional to that of the specific heat.

\begin{acknowledgments}
We are grateful to G.~D. Morris, B. Hitti, and D. Arsenau of the TRIUMF CMMS for assistance during the experiments, and to C.~M. Varma and K. Miyake for many useful discussions. This research was supported in part by the National Key Research and Development Program of China (Nos.~2017YFA0303104 and 2016YFA0300503), the National Natural Science Foundation of China No.~11474060, the development Fund Project of Science and Technology on Surface Physics and Chemistry Laboratory (Grant No.~XKFZ201602), the U.S. National Science Foundation under grant Nos.~DMR-1105380 (CSU-Los Angeles), DMR-1506677 (CSU-Fresno), and DMR-0802478 (UCSD). The research at UCSD was also supported by the U. S. Department of Energy grant No.~DE-FG02-04ER46105. This work was performed in part at the Aspen Center for Physics, which is supported by National Science Foundation grant PHY-1607611.
 \end{acknowledgments}

\raggedright

%

\end{document}